\documentclass[a4,prd,
nofootinbib,
preprintnumbers
,twocolumn
]{revtex4}
\usepackage{graphicx,epsf,epsfig}
\usepackage{amsmath, amssymb, amsfonts,bbm}
\usepackage{hyperref}
\usepackage[T1]{fontenc}
\usepackage[utf8]{inputenc}

\usepackage{color}

\newcommand{\be}{\begin{equation}}
\newcommand{\ee}{\end{equation}}
\newcommand{\bd}{\begin{displaymath}}
\newcommand{\ed}{\end{displaymath}}
\newcommand{\bea}{\begin{eqnarray}}
\newcommand{\eea}{\end{eqnarray}}
\newcommand{\nn}{\nonumber}

\begin{document}
\title{Renormalization group equations and matching in a general quantum field theory with kinetic mixing}
 \preprint{CFTP/13-018}
 \author{Renato M. Fonseca}\email{renato.fonseca@ist.utl.pt}
 \affiliation{Centro de F\'isica Te\'orica de Part\'iculas, Instituto Superior T\'ecnico,
 Universidade de Lisboa, Av. Rovisco Pais 1, 1049-001 Lisboa, Portugal}
 \author{Michal Malinsk\'y}\email{malinsky@ipnp.troja.mff.cuni.cz}
 \affiliation{Institute of Particle and Nuclear Physics, Faculty of Mathematics and Physics, Charles University in Prague, V Hole\v{s}ovi\v{c}k\'ach 2, 180 00 Praha 8, Czech Republic }
 \author{Florian Staub}\email{fnstaub@th.physik.uni-bonn.de}
 \affiliation{Bethe Center for Theoretical Physics \& Physikalisches Institut der, Universit\"at Bonn, Nu{\ss}allee 12, 53115 Bonn, Germany}
\begin{abstract}
We work out a set of simple rules for adopting the two-loop renormalization group equations of a generic gauge field theory given in the seminal works of Machacek and Vaughn to the most general case with an arbitrary number of Abelian gauge factors and comment on the extra subtleties possibly encountered upon matching a set of effective gauge theories in such a framework. 
\end{abstract}
\maketitle
\section{Introduction}
Since the advent of the renormalization group (RG) techniques in 
the early 1970's \cite{Gross:1973id,Gross:1973ju,Politzer:1973fx,Wilson:1973jj} the $\beta$-functions and the anomalous dimensions
of a wide range of theories have been calculated to a several loops. In this context, 
the series of 1980's papers by Machacek and Vaughn~\cite{Machacek:1983tz,Machacek:1983fi,Machacek:1984zw} in which the existing results have been wrapped up into a set of explicit two-loop formulae for (almost) any renormalizable gauge theory represents a cornerstone underpinning many recent precision calculations. As such, these results were subject to a thorough re-evaluation in the last two decades; remarkably enough, they passed with only very minor corrections, see, e.g., \cite{Luo:2002ti}.

Nevertheless, there is a class of scenarios that was left aside in the original works~\cite{Machacek:1983tz,Machacek:1983fi,Machacek:1984zw,Luo:2002ti}, in particular,  those theories that besides a semi-simple gauge structure feature more than a single Abelian gauge factor. 
Indeed, a qualitatively new feature -- a possible mixing among the gauge-invariant kinetic forms of the Abelian gauge fields~\cite{Holdom:1985ag,Babu:1997st} -- arises with more than a single $U(1)$ at play.

The impact of such a ``gauge kinetic mixing''
on the structure of the renormalization group equations (RGEs) and, subsequently, on the evaluation of the running gauge couplings has been studied in great detail in~\cite{delAguila:1988jz,delAguila:1987st}; similarly, the general two-loop $\beta$-functions for all 
other dimensionless parameters were derived in~\cite{Luo:2002iq}. 
However, up to our best knowledge, there are no results for {\em dimensionfull} parameters available in the existing literature. 

In this letter, we aim at filling this gap. In order to do so, we use the methods advocated in Ref.~\cite{Fonseca:2011vn} where the general two-loop results for  softly broken renormalizable supersymmetric
theories have been given. As we will show, this approach also significantly simplifies 
the derivation of the $\beta$-functions for the dimensionless parameters as 
compared to, e.g.,  Ref.~\cite{Luo:2002iq}.

In Sect.~\ref{sec:method} the method of Ref.~\cite{Fonseca:2011vn} is briefly reviewed. In Sect.~\ref{sec:results} we present a set of substitution rules  which admit for a straightforward generalization of the 
relevant formulae of
Ref.~\cite{Machacek:1983tz,Machacek:1983fi,Machacek:1984zw,Luo:2002ti} to the case with multiple $U(1)$ gauge factors and comment in brief on the subtleties that  may be encountered upon matching a set of effective gauge theories in such a framework. Then we conclude in Sect.~\ref{sec:conclusion}.

\section{The Method}
\label{sec:method}
The main novelty encountered in theories with more than a single $U(1)$ gauge factor is due to the new terms in Lagrangian
connecting the field strength tensors of different $U(1)$'s:
\be\label{eq:xi}
{\cal L}_{kin.}\ni -\tfrac{1}{4} F^T_{\mu\nu}\xi F^{\mu\nu}\,.
\ee
Here, the relevant field tensors have been grouped into an $n$-dimensional
vector $F_{\mu\nu}$ (with $n$ denoting the number of independent gauged $U(1)$'s) and $\xi$ is an $n\times n$ real and symmetric matrix.
This, in turn, introduces $\tfrac{1}{2}n(n-1)$ extra dynamical parameters.
In order to construct the RGEs for such a theory, one would have to (i)
derive the $\beta$-functions governing these extra parameters and (ii) 
modify the $\beta$-functions of all other parameters, such as the gauge 
or Yukawa couplings as well as the relevant anomalous dimensions. Such a  ``straightforward'' approach has been used in Ref.~\cite{Luo:2002iq}
to derive the two-loop $\beta$-functions for all dimensionless parameters of a generic (non-supersymmetric) renormalizable gauge theory. 

Alternatively, one can work in a renormalization scheme in which 
the $\xi$-parameter in Eq.~(\ref{eq:xi}) is absorbed by a suitable redefinition of the gauge fields, namely, 
\be
V\to \xi^{1/2}V\,.
\ee 
which, indeed, provides a canonically normalized gauge sector. However, such a redefinition impacts the non-derivative part of the original
 covariant derivative:
\be
Q_i^{T}{\tilde G} V \to Q_i^{T}{\tilde G}\xi^{-1/2}V\,,
\ee 
where $\tilde G$ is the original diagonal matrix\footnote{with indices in the
group and gauge-field spaces, respectively} of $n$ individual gauge couplings
associated to the $n$  Abelian gauge factors,  and  $Q_i$ is the
vector\footnote{with a lower index assigning the corresponding
matter-field} of the
relevant $U(1)$ charges. Similarly, the gauge-kinetic counterterm is transformed,
\be
Z_{V}^{1/2}\xi_{B}Z_{V}^{1/2}-\xi\to
\xi^{-1/2}Z_{V}^{1/2}\xi_{B}Z_{V}^{1/2}\xi^{-1/2}-1 \; \equiv \delta
Z_{\tilde{V}},
\ee
where the subscript $B$ denotes bare  quantities and $Z_{V}^{1/2}$ is the
original (diagonal) gauge-field renormalization factor $V_{B}=Z_{V}^{1/2}V$.
Hence, the $\xi^{-1/2}$ factor can  be ``hidden'' into a new set of  $\tfrac{1}{2}n(n-1)$ ``effective'' gauge couplings whose combinations populate off-diagonal entries of an ``extended gauge-coupling matrix'' 
\be\label{Gmatrix}
G\equiv \tilde G \xi^{-1/2}\,,
\ee 
and a suitably redefined gauge-kinetic counterterm.
In this scheme, the
gauge-kinetic counterterm $\delta Z_{\tilde{V}}$ 
is naturally off-diagonal and 
ready to absorb the ultraviolet divergences in the off-diagonal two-point functions.
Therefore, it is sufficient to work with a matrix-like gauge-coupling structure (\ref{Gmatrix}), 
forgetting entirely about the $\xi$-origin of its off-diagonal entries. 

Needless to say, these two strategies are entirely equivalent in theory; in practice, however, the latter is much more suitable for our task as it essentially amounts to
replacing all the  polynomials including individual gauge couplings in~\cite{Machacek:1983tz,Machacek:1983fi,Machacek:1984zw} by the relevant matrix structures, with no
need to deal with
the evolution equations for the $\xi$ matrix.

Let us also anticipate that the set of substitution rules given in the next section
is quite short and these rules are much easier to apply in practice than the results 
quoted in Ref.~\cite{Luo:2002iq}. In addition, they are entirely universal, i.e., they can be readily used to derive
the correct $\beta$-functions for all parameters in the theory,  not just the dimensionless ones. 
Finally, we note that due to the absence of gauginos and quartic couplings stemming from $D$-terms the current list is also significantly shorter than its supersymmetric counterpart given in Ref.~\cite{Fonseca:2011vn}.

\section{Results}
\label{sec:results}

In this section, we work out all the necessary substitution rules underpinning the general construction of the two-loop RGEs for 
an arbitrary non-supersymmetric renormalizable quantum field theory using the results of Ref.~\cite{Luo:2002ti}. In order to make the discussion more self-contained we shall also provide the generalization of the ``canonical'' Weinberg's~\cite{Weinberg:1980wa} and Hall's~\cite{Hall:1980kf} gauge theory matching formulae.
\subsection{The RGE structure}
Quantities related to the Abelian gauge factors appear in 
terms which explicitly depend on group generators,
as well as gauge group structure constants. Following
the convention of Ref.~\cite{Luo:2002ti}, the covariant derivatives for real scalar fields $\Phi_a$ 
and Weyl fermions $\Psi_j$ are written as\footnote{Repeated 
Greek indices, which refer to gauge bosons, are assumed to be summed over. 
Summations over all other types of indices are indicated explicitly.} 
\begin{align}
\label{eq:CovDer1}
D_\mu \Phi_a = & \partial_\mu \Phi_a - i g \sum_{b} \Theta_{ab}^\alpha V_\mu^\alpha \Phi_b\,, \\
\label{eq:CovDer2}
D_\mu \Psi_j = & \partial_\mu \Psi_j - i g \sum_{k} t^\alpha_{jk} V_\mu^\alpha \Psi_k\,.
\end{align}
For non-Abelian groups, $\Theta^\alpha$ and $t^\alpha$ are hermitian matrices. In addition, $\Theta^\alpha$'s
are purely imaginary and antisymmetric. The quadratic Dynkin and Casimir indexes are defined as
\begin{align}
& C^{ab}(S) = \left(\Theta^{\alpha}\Theta^{\alpha}\right)_{ab}, \,\, S(S)\delta^{\alpha\beta} = \mbox{Tr}(\Theta^\alpha \Theta^\beta)\,, \\
& C^{ab}(F) = \left(t^{\alpha}t^{\alpha}\right)_{ab}, \,\, S(F)\delta^{\alpha\beta} = \mbox{Tr}(t^\alpha t^\beta)\,. 
\end{align}
The $\beta$-functions and anomalous dimensions for a general non-supersymmetric renormalizable quantum field theory including an arbitrary number of simple gauge factors but at most one $U(1)$ are obtained from
the formulas and
substitution rules given in Refs.~\cite{Machacek:1983tz,Machacek:1983fi,Machacek:1984zw,Luo:2002ti}. 
In what follows, we shall generalize these
substitution rules to cover the case with an arbitrary number of Abelian factors.

The gauge group is taken to be ${\cal G}_{A}\otimes {\cal G}_{B}\otimes \ldots\otimes U\left(1\right)^{n}$,
where the ${\cal G}_{X}$'s are simple groups (uppercase indices 
shall be used only for simple factors). As mentioned before, the $U(1)$ sector should be treated as a whole and described in terms of a general real 
$n\times n$ gauge-coupling matrix $G$ and a column vector
of charges $Q^S_{i}$ for each scalar $\Phi_{i}$ and $Q^F_{i}$ for each fermion $\Psi_i$. 
Using this convention, we can rewrite Eqs.~(\ref{eq:CovDer1}) and (\ref{eq:CovDer2}) as
\begin{align}
\label{eq:CovDer1new}
D_\mu \Phi_a = & \partial_\mu \Phi_a - i \tilde{\delta}_{ab} \left(Q_{b}^{S}\right)^{T} 
  G V \Phi_b - i \sum_{A,b} g_A \Theta_{A,ab}^\alpha V_\mu^\alpha \Phi_b\,, \\
\label{eq:CovDer2new}
D_\mu \Psi_j = & \partial_\mu \Psi_j - i \delta_{jk} \left(Q_{k}^{F}\right)^{T}
  G V \Psi_k - i \sum_{A,b} g_A t^\alpha_{A,jk} V_\mu^\alpha \Psi_k\,,
\end{align}
where $V$ is a vector containing the $U(1)$ gauge bosons. In passing we introduced an antisymmetric tensor $\tilde{\delta}_{ab}$ equal to the imaginary
 unit $i$ if $a$ and $b$ are the real and imaginary components of an eigenstate of the $U(1)$ gauge interactions, respectively; otherwise, $\tilde{\delta}_{ab}=0$. 

We also note that the forms
\begin{equation}
W^R_{i}\equiv G^{T}Q^R_{i} 
\end{equation}
where $i$ numbers the scalar ($R=S$) or fermion ($R=F$) of the theory are the only
combinations of $Q^R_{i}$ and $G$ which appear in the Lagrangian and, hence, the Abelian charges should appear in the relevant RGEs only through these quantities. 

The only non-trivial substitution rule to be applied at the one-loop level for the $\beta$-function of the gauge couplings reads:
\begin{align}
g^3 S(R) & \rightarrow G \sum_{p}W^{R}_p \left(W^{R}_p\right)^{T}\,. 
\end{align}
Here, as well as below, $\sum_p$ runs over all fermions for $R=F$, or scalars for $R=S$. 

At two loops, the substitution rules for the matrix $G$ of the Abelian gauge couplings read:
\begin{align}
\nonumber g^{5}C(R)S(R)&\rightarrow\sum_{p}GW_{p}^{R}\left(W_{p}^{R}\right)^{T} \\
&\times\left[\sum_{B}g_{B}^{2}C_{B}^{pp}(R)+\left(W_{p}^{R}\right)^{T}W_{p}^{R}\right]\,, \\
g^3 C^{ab}(R)/d({\cal G}) & \rightarrow   \delta_{ab}G W^R_{a} \left(W^R_{a}\right)^{T}\,. 
\end{align}
Note that all terms involving $C({\cal G})$ vanish and, hence, the corresponding formulae remain intact. For a non-Abelian group factor ${\cal G}_{A}$, only one expression needs to be modified:
\begin{align}
\nonumber g^{5}C(R)S(R)\rightarrow & g_{A}^{3}\sum_{p}C_{A}^{pp}(R)/d\left({\cal G}_{A}\right)\\
 & \times\left[\sum_{B}g_{B}^{2}C_{B}^{pp}(R)+\left(W_{p}^{R}\right)^{T}W_{p}^{R}\right]\,.
\end{align}
For the calculation of the $\beta$-functions for the parameters appearing in the potential, one needs
\begin{widetext}
\begin{align}
g^{2}C^{ab}(R) & \rightarrow \sum_{A}g_{A}^{2}C^{ab}_A(R)+ \delta_{ab} \left(W_{a}^{R}\right)^{T} W^R_{a}\,,\\
g^{4}C^{ab}(R)C^{cd}(R')&\rightarrow\left[\sum_{A}g_{A}^{2}C_{A}^{ab}(R)+\delta_{ab}\left(W_{a}^{R}\right)^{T}
W_{a}^{R}\right]\left[\sum_{B}g_{B}^{2}C_{B}^{cd}(R')+\delta_{cd}\left(W_{c}^{R'}\right)^{T}W_{c}^{R'}\right]\,, \\
g^4 S(R) C^{ab}(R') & \rightarrow \sum_{A}g_{A}^{4}S_A(R)C_A^{ab}(R')+\delta_{ab} \sum_{p}\left[\left(W_{p}^{R}\right)^{T}W^{R'}_{a}\right]^{2}\,.
\end{align}
\end{widetext}
In addition, the gauge charges (i.e., generators) appear explicitly in the $\beta$-functions. The corresponding replacement rules follow from the generalized form of the covariant derivatives given in Eqs.~(\ref{eq:CovDer1new}) and (\ref{eq:CovDer2new}). For terms including fermionic generators, they read
\begin{widetext}
\begin{align}
g^2 t^\alpha_{ij} t^\alpha_{kl} & \rightarrow \sum_A g_A^2 t^\alpha_{A,ij} t^\alpha_{A,kl} + \delta_{ij} \delta_{kl} \left(W^F_i\right)^T W^F_k\,, \\
g^{4}\{\Theta^{\alpha},\Theta^{\beta}\}_{ab}t_{ij}^{\alpha}t_{kl}^{\beta} & \rightarrow\sum_{A}\sum_{B}g_{A}^{2}g_{B}^{2}
\left\{ \Theta_{A}^{\alpha},\Theta_{B}^{\beta}\right\} _{ab}t_{A,ij}^{\alpha}t_{B,kl}^{\beta}+2\delta_{ab}\delta_{ij}
\delta_{kl}\left(W_{a}^{S}\right)^{T}W_{i}^{F}\left(W_{b}^{S}\right)^{T}W_{k}^{F} \nonumber \\
 & \hspace{0.3cm}+\sum_{A,p}g_{A}^{2}\left[\tilde{\delta}_{ap}\left(W_{a}^{S}\right)^{T}\Theta_{A,pb}^{\alpha}+\tilde{\delta}_{pb}
 \left(W_{b}^{S}\right)^{T}\Theta_{A,ap}^{\alpha}\right]\left(\delta_{ij}W_{i}^{F}t_{A,kl}^{\alpha}+
 \delta_{kl}W_{k}^{F}t_{A,ij}^{\alpha}\right)\,,
\end{align}
and, similarly, for terms involving $t^*$. In order to generalize the factors involving scalar generators only,
we introduce the following shorthand notation: 
\begin{align}
\widetilde{\Lambda}_{ab,cd} & \equiv\sum_{A}g_{A}^{2}\left(\Theta_{A}^{\alpha}\right)_{ac}\left(\Theta_{A}^{\alpha}\right)_{bd}+\tilde{\delta}_{ac}\tilde{\delta}_{bd}\left(W_{a}^{S}\right)^{T}W_{b}^{S}\,,\\
\widetilde{\Lambda}_{ab,cd}^{S} & \equiv\sum_{A}g_{A}^{4}S_{A}\left(R\right)\left(\Theta_{A}^{\alpha}\right)_{ac}\left(\Theta_{A}^{\alpha}\right)_{bd}+\tilde{\delta}_{ac}\tilde{\delta}_{bd}\sum_{p}\left(W_{a}^{S}\right)^{T}W_{p}^{S}\left(W_{p}^{S}\right)^{T}W_{b}^{S}\,,\\
\widetilde{\Lambda}_{ab,cd}^{G} & \equiv\sum_{A}g_{A}^{4}C\left({\cal G}_{A}\right)\left(\Theta_{A}^{\alpha}\right)_{ac}\left(\Theta_{A}^{\alpha}\right)_{bd}\,.
\end{align}
With this at hand, one obtains
\begin{align}
g^{2}\left(\Theta_{A}^{\alpha}\right)_{ac}\left(\Theta_{A}^{\alpha}\right)_{bd} & \rightarrow
\widetilde{\Lambda}_{ab,cd}\,,\\
g^{4}\left\{ \Theta^{\alpha},\Theta^{\beta}\right\} _{ab}\left\{ \Theta^{\alpha},\Theta^{\beta}\right\} _{cd} 
& \rightarrow 2\sum_{e,f}\left(\widetilde{\Lambda}_{ac,ef}\widetilde{\Lambda}_{ef,bd}+\widetilde{\Lambda}_{af,ed}\widetilde{\Lambda}_{ec,bf}\right)\,,\\
g^{6}C\left({\cal G}\right)\left\{ \Theta^{\alpha},\Theta^{\beta}\right\} _{ab}\left\{ \Theta^{\alpha},\Theta^{\beta}\right\} _{cd}
 & \rightarrow2\sum_{e,f}\left(\widetilde{\Lambda}_{ac,ef}^{G}\widetilde{\Lambda}_{ef,bd}+\widetilde{\Lambda}_{af,ed}^{G}
 \widetilde{\Lambda}_{ec,bf}\right)\,,\\
g^{6}S\left(R\right)\left\{ \Theta^{\alpha},\Theta^{\beta}\right\} _{ab}\left\{ \Theta^{\alpha},\Theta^{\beta}\right\} _{cd}
 & \rightarrow2\sum_{e,f}\left(\widetilde{\Lambda}_{ac,ef}^{S}\widetilde{\Lambda}_{ef,bd}+\widetilde{\Lambda}_{af,ed}^{S}
 \widetilde{\Lambda}_{ec,bf}\right)\,,\\
g^{6}\sum_{i}C\left(i\right)\left\{ \Theta^{\alpha},\Theta^{\beta}\right\} _{ab}\left\{ \Theta^{\alpha},\Theta^{\beta}\right\} _{cd}
 & \rightarrow2\sum_{e,f}\sum_{\textrm{free }i}\left[\sum_{B}g_{B}^{2}C_{B}\left(i\right)+\left(W_{i}^{S}\right)^{T}W_{i}^{S}\right]
 \left(\widetilde{\Lambda}_{ac,ef}\widetilde{\Lambda}_{ef,bd}+\widetilde{\Lambda}_{af,ed}\widetilde{\Lambda}_{ec,bf}\right)\,,
\end{align}
where $C_A\left(i\right)$ is the same as $C_A^{ii}(S)$ (see Refs.~\cite{Machacek:1983tz,
Machacek:1983fi,Machacek:1984zw,Luo:2002ti}). Note also that, unlike $e$, $f$ or $p$ in previous expressions, the index $i$ is to be summed over the free indices only  (not over all scalars): $i=a,b,c,d$, $i=a,b,c$ or $i=a,b$ depending on the specific setting.
\end{widetext}

\subsection{Matching conditions}
Subtleties related to the kinetic mixing are also encountered in the matching formulae relating two sets of gauge couplings parameterizing the high- and low-energy effective theories upon passing through a characteristic scale where heavy degrees of freedom are integrated out. 

For the sake simplicity, let us start with the basic formulae~\cite{Weinberg:1980wa,Hall:1980kf} relating a gauge coupling $g$ of a simple gauge group ${\cal G}$ (associated to some grand unified theory) to the effective couplings $g_{C}$ associated to the set of preserved gauge factors ${\cal G}_{C}$ relevant to its broken-phase description below a characteristic scale $\mu$. These obey 
\be\label{nonabeliansimple}
g_{C}^{-2} = g^{-2} - \lambda_C(\mu)\,,
\ee
where 
\bea\label{lambda}
\lambda_{C}(\mu)&=&\frac{1}{48\pi^{2}}\Delta_{C}+\frac{1}{8\pi^{2}}\left[-\frac{11}{3}S_{C}(V_{C})\log\frac{M_{V_{C}}}{\mu}\right. \\
&+&\left.\frac{4}{3}\kappa_{F_{i}}S_{C}(F_{i})\log\frac{M_{F_{i}}}{\mu}+\frac{1}{3}\eta_{S_{b}}S_{C}(S_{b})\log\frac{M_{S_{b}}}{\mu}\right]\; , \nn
\eea
is the ``threshold'' function in the $\overline{\rm MS}$ scheme (and in dimensional regularization); $\Delta_{C}=C({\cal G})-C({\cal G}_{i})$ is the difference of the two group Casimirs (i.e., the indexes of the adjoint representations of ${\cal G}$ and ${\cal G}_{C}$), $S_{C}$ denote the Dynkin indexes of the fermions $F_{i}$  and scalars $S_{b}$ (with masses $M_{F_{i}}$ and $M_{S_{b}}$) with respect to ${\cal G}_{C}$, $\kappa_{F_{i}}=1$ if $F_{i}$ is a Dirac spinor and $\tfrac{1}{2}$ for Weyl, and $\eta_{S_{b}}=1$ if  $S_{b}$ is a complex scalar and $\tfrac{1}{2}$ for real. Note that the (Feynman gauge) Goldstone boson's effects have been conveniently included into the last term among those of general massive scalar fields; similarly, the effects of ghosts (segregated in the original works~\cite{Weinberg:1980wa,Hall:1980kf}) have been subsumed into the first (gauge) term in the bracket.   

Note also that in this convention the coefficients of the $\mu$-dependent terms in the bracket of Eq.~(\ref{lambda}) closely resemble those popping up in the classical one-loop $\beta$-function formula for non-Abelian gauge theories. This, indeed, is crucial to ensure the leading-order matching-scale independence of the effective gauge couplings at two loops. 

Needless to say, the same should happen in the case when there is more than a single $U(1)$ gauge factor among either the effective high- or low-energy gauge factors (or both).  Then, however, the $\beta$-functions are conveniently written in terms of the gauge-coupling matrix structure $G$, see Eq.~(\ref{Gmatrix}), so one may expect similar changes in the relevant threshold functions $\lambda(\mu)$.

To this end, let us focus solely onto the matching among two gauge theories with $U(1)^{n}$ and $U(1)^{n'}$ gauge groups ($n'<n$), respectively\footnote{An interested reader can find a more complete account of the general case including extra sets of non-Abelian factors in the recent study~\cite{Bertolini:2013vta}.}. Without loss of generality one may assume that their generators fulfil    
\be
Q'=P Q\,,
\ee 
where $P$ is an $n'\times n$ matrix projector onto the (primed) subspace of the conserved charges obeying $PP^{T}=\mathbbm{1}'$. 
The relevant gauge-coupling matrices $G'$ and $G$ are then related by the formula
\be\label{matching}
(G'G'^{T})^{-1}=P[(GG^{T})^{-1}-\Lambda(\mu)]P^{T}\,,
\ee
where the ``matching function'' $\Lambda(\mu)$ reads at the leading (i.e., one-loop) order 
\bea\label{Lambda}
\Lambda(\mu)& =& \frac{1}{8\pi^{2}}G^{T-1}\left[\frac{4}{3}\sum_{F_{i}}\kappa_{F_{i}} W^{F}_{i}(W_{i}^{F})^{T}\log\frac{M_{F_{i}}}{\mu}\right.\nn \\& &+ \left.\frac{1}{3}\sum_{S_{b}}\eta_{S_{b}}W^{S}_{b}(W^{S}_{b})^{T}\log\frac{M_{S_{b}}}{\mu}\right]G^{-1}\,.
\eea
As before, $F_{i}$ and $S_{b}$ denote the fermions and scalars in the theory, respectively, and there are no contributions similar to the first two terms in Eq.~(\ref{nonabeliansimple}) in the purely Abelian case. 
Note also that the transformation properties of $G^{T-1}WW^{T}G^{-1}$ under the change of bases in the charge and gauge-field spaces are identical to those of $(GG^{T})^{-1}$; hence, the algebraic structure of Eq.~(\ref{Lambda}) is fully justified.

There is perhaps one more point worth making here. The matrix structure of Eqs.~(\ref{matching}) and (\ref{Lambda}) is very important also for the self-consistency of the effective theory which, as long as the gauge interactions are considered, should care only about the mass and the ``effective'' charges of any given multiplet without any reference to its full-theory origin. 

Let us exemplify this on the specific case of the Goldstone boson with the Standard Model quantum numbers $(3,1,+\tfrac{2}{3})$ which provides the longitudinal component to the massive vector boson in a class of Pati-Salam extensions of the Standard Model (recall that there is indeed the $(3,1,+\tfrac{2}{3})$+h.c. vector among the gauge fields of $SU(4)_{C}\otimes SU(2)_{L}\otimes U(1)_{R}\equiv 421$ and we chose $U(1)_{R}$ instead of the full $SU(2)_{R}$ just to get rid of the extra vectors/Goldstones present in the LR-symmetric settings). Such a field typically emerges as a massless  mixture of several components with different 421 origin that, besides their obvious $SU(3)_{c}\times SU(2)_{L}$ charges can be conveniently classified in terms of the two independent Cartans $Q_{R}$ of $U(1)_{R}$ and (the physically normalized) $Q_{BL}$ of the $U(1)_{BL}$ subgroup of $SU(4)_{C}$. More specifically, let us consider $(3,1,+\tfrac{2}{3})$ originating from the mixture of $(3,1,0,+\tfrac{4}{3})\equiv S_{1}$ and $(3,1,+1,-\tfrac{2}{3})\equiv S_{2}$ encountered in the recent work~\cite{Bertolini:2013vta} and check that the threshold correction to the effective $U(1)_{Y}$ gauge coupling due to $(3,1,+\tfrac{2}{3})$+h.c. being integrated out does not depend on the mixing among $S_{1}$ and $S_{2}$, although the relevant $W_{b}^{S}(W_{b}^{S})^{T}$ matrices in (\ref{Lambda}) are obviously different:
\bea
\sum_{b}W_{b}^{S_{1}}(W_{b}^{S_{1}})^{T}& =& (1\;\sqrt{\tfrac{3}{8}})
\left(\begin{array}{cc}
0 & 0 \\
0 & +\tfrac{16}{3}
\end{array}\right)
\left(\begin{array}{c}
1 \\
\sqrt{\tfrac{3}{8}}
\end{array}\right)\,,\nn\\
\sum_{b}W_{b}^{S_{2}}(W_{b}^{S_{2}})^{T}& =& (1\;\sqrt{\tfrac{3}{8}})
\left(\begin{array}{cc}
+3 & -2 \\
-2 & +\tfrac{4}{3}
\end{array}\right)
\left(\begin{array}{c}
1 \\
\sqrt{\tfrac{3}{8}}
\end{array}\right)\nn\,.
\eea
Indeed, sandwiching these structures among the projectors onto the SM hypercharge $P_{Y}=(\sqrt{\tfrac{3}{5}}\;\sqrt{\tfrac{2}{5}})$ as suggested by formula (\ref{matching}) produces the same result for both $S_{1}$ and $S_{2}$, or any mixture of the two fields.

\section{Conclusion}
\label{sec:conclusion}
In this letter we have worked out the rules for adopting 
the two-loop renormalization group equations given in the seminal works~\cite{Machacek:1983tz,Machacek:1983fi,Machacek:1984zw} (in the notation of \cite{Luo:2002ti}) to the most general case of a non-supersymmetric renormalizable gauge theory featuring an arbitrary number of Abelian gauge factors. It is important to stress that our method is entirely universal as it can be readily used to derive
the $\beta$-functions for {\em all} parameters in the theory,  not just the dimensionless ones considered previously in the literature. The approach follows closely which we used formerly in the supersymmetric case~\cite{Fonseca:2011vn} with just minor modifications reflecting a slightly different convention used in the primary papers. We also comment on the extra subtleties  often encountered upon matching a set of effective gauge theories in such a framework. 
\section*{Acknowledgements}
We thank Werner Porod for fruitful discussions and his remarks concerning the manuscript. 
We thank Florian Lyonnet for pointing out wrong powers of $g$ in eqs.~(23), (24), (27) and (28). 
The work of M.M. is supported by the Marie-Curie Career Integration Grant within the 7th European Community Framework Programme
FP7-PEOPLE-2011-CIG, contract number PCIG10-GA-2011-303565 and by the Research proposal MSM0021620859 of the Ministry of Education, Youth and Sports of the Czech Republic. 
The work of R.M.F has been supported by \textit{Fundação para a Ciência
e a Tecnologia} through the fellowship SFRH/BD/47795/2008. He also
acknowledges the financial support from grants CFTP-FCT UNIT 777,
CERN/FP/123580/2011 and PTDC/FIS/102120/2008.
\bibliographystyle{h-physrev5}
\bibliography{bib}
\end{document}